# Adaptive Self-Improvement for Smarter Energy Systems using Agentic Policy Search

ALEXANDER SOMMER, Friedrich-Alexander-Universität Erlangen-Nürnberg, Germany
PETER BAZAN, Friedrich-Alexander-Universität Erlangen-Nürnberg, Germany
BEHNAM BABAEIAN, Friedrich-Alexander-Universität Erlangen-Nürnberg, Germany
JONATHAN FELLERER, Friedrich-Alexander-Universität Erlangen-Nürnberg, Germany
WARREN B. POWELL, Princeton University, USA
REINHARD GERMAN, Friedrich-Alexander-Universität Erlangen-Nürnberg, Germany

Controlling energy systems usually involves manually designed policies for decision-making, which can be complex and time-consuming to develop. This process requires interdisciplinary collaboration among multiple domain experts, resulting in slow and inflexible adaptation to rapidly changing environments. Large Language Models (LLMs) offer a promising paradigm shift by integrating extensive contextual knowledge with the capability to generate structured, executable code.

We present *Agentic Policy Search* (APS)—a novel hierarchical optimization framework in which LLMs act as autonomous agents that propose complete control logics, translate them into executable code, and iteratively improve them through direct system feedback. We apply APS to a residential energy system with PV, battery, demand, and dynamic electricity prices. Within just seven simulated days, the method yields a net profit of up to 6.20 € compared to the no-battery reference scenario (−10.70 €), nearly matching the global optimum of a perfectly informed linear program. By combining LLM-driven policy search with the generation of human-interpretable control logic, APS effectively bridges adaptability and traceability in energy management—while also offering a transferable framework for agentic optimization in other domains.

## 1 Introduction

Modern energy systems are characterized by increasing complexity due to the growth of distributed generation, storage technologies, and dynamic consumption patterns. Effective management of these interconnected components requires sophisticated control methods and adaptive mechanisms.

Traditionally, decision-making in energy systems has relied on predefined functions, known as policies, which specify appropriate actions depending on the current system state. The development and refinement of these policies typically depend on specialized expertise, demanding extensive domain knowledge across different modeling paradigms and optimization techniques.

However, the accelerating pace of technological advancements, regulatory shifts, and evolving market conditions continuously reshapes energy systems. This rapid evolution poses significant challenges: policies optimized for past conditions quickly become suboptimal, limiting operational efficiency and flexibility. Moreover, reliance on centralized expertise slows down the responsiveness necessary for timely policy adaptation, hindering both short-term operations and long-term strategic planning. As a result, opportunities for system-wide coordination—such as sector coupling—and efficiency enhancements remain underexploited.

To address these challenges, LLMs are gaining traction in energy systems research. Their ability to access broad knowledge and generate context-aware insights helps democratize expertise that was once exclusive to specialized fields. By making this knowledge more widely accessible, LLMs enable interdisciplinary collaboration and support solutions that previously required large teams of domain experts, accelerating informed and adaptive decision-making in complex energy environments.

Building on this perspective, this paper examines how LLMs can move toward active involvement in policy development and iterative system improvement. We focus in particular on their potential to autonomously implement and assess control strategies, using system feedback to continuously search for, test, and refine solutions that enhance overall performance

## 2 Related Work

LLMs have predominantly been applied within energy system simulations [5, 50] and as supportive expert advisors in higher-level strategic decision-making [9]. Nevertheless, their direct application as controllers in energy management remains constrained. Particularly when faced with tasks demanding advanced algorithmic reasoning, existing models frequently fail to produce solutions that are both robust and broadly applicable [39].

While approaches such as chain-of-thought prompting [46] can improve performance, LLMs remain inefficient at scaling for high-dimensional problems and lack formally verifiable correctness guarantees [13, 49]. On the other hand, LLMs excel at converting optimization tasks into software code, as demonstrated across HVAC control, electrical vehicle charging, and power systems [16, 19]. Recent advances show how LLMs can formulate solutions as abstract hypotheses before programmatic implementation and verification [43], generate adaptable software code for decision-making tasks [3, 24], and employ agent-driven structures for precise strategy generation [18] as well as algorithmic discovery [25] through collaborative reasoning.

In the technical control domain, Guo et al. [12] and Zahedifar et al. [52] demonstrate how LLMs can support classical controllers by tuning parameters such as gains in PID or Lyapunov-based schemes. However, their approaches operate within fixed control architectures and leave the underlying policy structure unchanged. Ishida et al. [14] propose *LangProp*, a framework in which an LLM iteratively improves controller code through a prompt-based evolution loop. This results in PID-like controllers for CartPole and simple driving scenarios. Bosio and Müller [1] follow a similar evolutionary paradigm, evolving interpretable code for classic benchmark tasks like pendulum swing-up and ball-in-cup. Their prompt logic, however, is handcrafted and based on static templates, limiting generalization. Cui et al. [4] take a modular approach by orchestrating a multi-agent LLM architecture, where specialized agents sequentially handle model selection, algorithm synthesis, tuning, and validation for power electronics controllers. Liu et al.'s *RePower* [21] presents a fully autonomous research platform that uses a single LLM in a two-stage reflection-and-evolution loop, supported by an error-check agent, to generate and refine solution functions for typical offline power systems tasks. While effective across supervised-learning benchmarks, its design remains batch-oriented; although hardware integration is supported, the framework does not incorporate any mechanism for continuous online adaptation during operation.

In contrast, our hierarchical **Agentic Policy Search (APS)** framework bridges both offline exploration and online adaptation. It enables continuous self-improvement of executable control policies during real or simulated operation, forming a closed-loop architecture suited for dynamic and evolving energy environments.

## 3 Contribution

We introduce a mathematical framework for the adaptive self-improvement of control policies in smart energy systems. Our core contribution is a hierarchical decision architecture that integrates LLMs into a stochastic policy search, thereby unifying concepts from stochastic optimization and machine learning in a novel way. Specifically, we first propose the theoretical formulation in Section 5 of a three-level design:

(1) a **control policy level** (Section 5.2) that makes operational decisions within the energy system,
(2) a **policy generation level** (Section 5.3) that uses an LLM to synthesize executable control logic, and
(3) a **meta-policy level** (Section 5.4) that steers iterative improvement based on performance feedback.

Based on this, we present the corresponding practical implementation in Section 6.

Unlike static code generation approaches, our APS-framework enables continuous adaptation through environmental interaction, addressing key shortcomings of LLM-based controllers such as limited traceability, reproducibility, and robustness in dynamic settings.

We demonstrate the approach in a residential energy system scenario featuring battery storage and renewable generation. The agent achieves profits close to the finite-horizon optimum, despite lacking prior knowledge of external inputs. While developed in the energy domain, the architecture is *general and extensible* to a wide range of sequential decision-making problems.

## 4 Fundamentals: Universal Modeling Framework

This work employs the universal modeling framework by Powell [28–30] as the basis for LLM-driven policy search. We begin by outlining the theoretical foundations in the following section.

The core objective is to find an optimal policy specification $\pi \in \Pi$ that solves the following problem:

$$\max_{\pi} \mathbb{E}\left\{ \sum_{t=0}^{T} C_t(S_t, X_t^{\pi}(S_t), W_{t+1}) \mid S_0 \right\}, \tag{1}$$

where the system dynamics are described by the transition equation

$$S_{t+1} = S^M(S_t, x_t, W_{t+1}). \tag{2}$$

The variables and functions are defined as follows:

- $S_t$: the state variable containing all relevant system information at time $t$,
- $X_t^{\pi}(S_t)$: the decision function - called policy - that maps each state to a decision variable $x_t \in \mathcal{X}_t$ with $\pi = \{f, \theta\}$ where $f \in \mathcal{F}$ is the function's type and $\theta \in \Theta^f$ are the corresponding parameters.
- $W_{t+1}$: the exogenous information available at time $t+1$, which may depend on the current state $S_t$ and/or the decision $x_t$,
- $S^M(\cdot)$: the transition function describing the temporal evolution of the system, where $M$ is a label indicating the model, and
- $C_t(\cdot)$: the objective function evaluating the contribution or costs of the chosen decision.

Each policy $X^{\pi}$ is defined by a specification $\pi$, which comprises a function type $f$ (e.g., an analytical function) and a set of tunable parameters $\theta$, such as weights of a neural network. Our objective is to identify a policy specification that yields a sequence of decisions $\{x_0, x_1, \ldots, x_T\}$ maximizing the expected cumulative reward (or minimizing cost) over a planning horizon $T$. For this search over function classes the universal modeling framework [29] identifies four fundamental classes of policies that encompass any method for making decisions:

(1) **Policy function approximations (PFAs)**: These are analytical functions that map a state directly to an action without solving an embedded optimization problem. Examples range from simple, rule-based approaches to complex functions such as neural networks.
(2) **Cost function approximations (CFAs)**: This involves solving a parameterized optimization problem—usually a simplified version of the original problem—where parameters are tuned to yield policies that perform well on average over time.
(3) **Value function approximations (VFAs)**: These policies are based on the Bellman equation. They make a decision by optimizing the immediate reward and an approximation of the future value of the subsequent state. Classic examples are methods of approximate dynamic programming, such as Q-learning.
(4) **Direct lookahead approximations (DLAs)**: These policies make a decision in the here and now by solving an explicit model of the future over a specific planning horizon. The lookahead model itself can be deterministic (e.g. Model Predictive Control, MPC) or stochastic (e.g. Monte Carlo Tree Search, MCTS).

These four classes are universal, in that *any* method for making decisions falls into one of these four classes, or a hybrid of multiple classes. Powell and Meisel [31] demonstrate, using an energy storage problem, that policies from each of the four classes (plus a hybrid) can perform best depending on the characteristics of the available information. In the following section, we show how LLMs can actively traverse the policy space $\Pi$ to identify and refine optimal decision strategies within our APS framework.

## 5 Theory: LLM-driven Policy Search via APS

Building on Powell's universal modeling framework [28–30] introduced in Section 4, we cast our new APS framework as an optimization problem for an energy system with nested decision levels, detailing each level and clarifying their interdependencies.

### 5.1 Overview: Hierarchical Decision Making

The proposed framework comprises three tightly interlinked sequential decision-making processes operating on different time scales:

(1) the **policy execution on control level** that directly operates the physical asset,
(2) a **policy implementation on generation level** that synthesises new control policies in code, and
(3) a **policy search on meta level** that governs the overall learning process.

During the strategic phase (Level 3), the objective is to establish a high-level control strategy, such as selecting a MPC approach, that yields into minimizing, e.g., the total cost in the energy system. In the subsequent operational phase (Level 2), this strategic decision is concretized into executable control software, ensuring compliance with relevant constraints and operational requirements. The deployed policy is then evaluated in the environment (Level 1), and the resulting insights are used to iteratively refine the strategy, progressively converging toward an optimal solution.

The next section details the execution layer (Level 1), where we examine how a given policy interacts with the environment. We then proceed to discuss the operationalization of abstract strategies into concrete code (Level 2), before turning to the strategic selection and refinement of policies at the meta level (Level 3).

### 5.2 Level 1: Control Policy Execution

At this level, the control policy is deployed to manage the actual physical or simulated system. The focus lies on decision execution, direct feedback collection, and immediate adaptation to operational conditions. In the following, we map elements of an energy system to the universal modeling framework introduced in Section 4. Variables associated with the energy system are grouped under *Canonical Elements* and are consistently denoted with the superscript es to indicate their relation to the energy system

*5.2.1 Canonical Elements.* The techno-economic state of the energy system at time $t$ is represented by the state vector $S_t^{\text{es}}$, which may include, for example, the state of charge of electrical storage units, temperatures of thermal storage, and estimated information about the future, such as market- or weather forecasts. Here, $t$ can represent either the time in reality or that of the simulation environment.

The operational decision $x_t^{\text{es}}$ is determined by a control policy $X_t^{\pi^{\text{es}}}$, defined as the mapping

$$x_t^{\text{es}} = X_t^{\pi^{\text{es}}}(S_t^{\text{es}}),$$

where $\pi^{\text{es}}$ contains the information about the control policy with its function type $f^{\text{es}} \in \mathcal{F}^{\text{es}}$ and tunable parameters $\theta^{\text{es}} \in \Theta^{\text{es}}$. In this context $X_t^{\pi^{\text{es}}}(S_t^{\text{es}})$ defines which action should be taken given the current energy system state, such as charging/discharging power, generator dispatch, or demand response metrics.

The impact of the decision is subject to the system dynamics $S^{M,\text{es}}$ describing the temporal evolution of the energy system state by

$$S_{t+1}^{\text{es}} = S^{M,\text{es}}(S_t^{\text{es}}, x_t^{\text{es}}, W_{t+1}^{\text{es}}). \tag{3}$$

These dynamics can reflect physical phenomena (e.g., energy flows through transmission lines) or information-driven processes (e.g., data processing or communication delays). Exogenous variables $W_{t+1}^{\text{es}}$ refer to information that is not available at the time $t$ of decision-making and whose realization becomes known only afterwards. This encompasses, for example, environmental feedback and measured values, updating state information or serve as the basis for calculating further performance metrics of the executed control policy.

*5.2.2 Objective Function.* Having introduced the canonical elements of the control layer, our goal is to find a control policy specification $\pi^{\text{es}}$ that generates a sequence of actions $x_t^{es}$ maximizing the following objective function $C_t^{\text{es}}$:

$$\max_{\pi^{\text{es}} \in \Pi^{\text{es}}} \mathbb{E}\left\{\sum_{t=0}^{T} C_t^{\text{es}}(S_t^{\text{es}}, X_t^{\pi^{\text{es}}}(S_t^{\text{es}}), W_{t+1}^{\text{es}}) \mid S_0^{\text{es}}\right\}.$$

On the control level, the optimization objective typically consists of maximizing a techno-economic value over the planning horizon $T$, such as:

- minimizing operational costs,
- maximizing market revenues, or
- reducing $CO_2$ emissions,

while adhering to technical, regulatory, and physical constraints, starting from the initial state $S_0^{\text{es}}$ of the energy system.

We adopt Powell's notation [28] for online optimization problems, which enables us to not only address the one-time derivation of a control policy, but also its continuous refinement. For example, in the event that consumption patterns change and the old control policy no longer works as planned.

Having defined the optimization objective, we now turn to the question of how to create such control policy with an LLM.

## 5.3 Level 2: LLM-driven Policy Implementation

At this intermediate layer, high-level control strategies are translated into executable policies. The challenge lies in ensuring that abstract policy concepts—such as those selected at the strategic layer—are expressed in concrete, operational code. This involves compiling domain knowledge, system constraints, and optimization logic into structured control programs suitable for deployment. For example, a decision to apply rule-based control can be implemented in Python as if-else logic that switches a battery on or off depending on current price signals.

*Canonical Elements.* We conceptualize an executable control policy as program code consisting of individual tokens. A *token* is defined in the literature as a basic textual unit, such as words, subwords, or specific source code elements (e.g., delimiters, parentheses) [36, 40]. For instance, the Python statement def policy(state): return decision would be tokenized into:

["def", " policy", "(state", "):",
" return", " decision"].

The autoregressive prediction of such token sequences enables LLMs to generate coherent and syntactically correct program code.

Our objective is to utilize this mechanism to translate a high-level policy description into a structured sequence of tokens that embodies a complete control policy implementation. Here, the central optimization challenge involves accurately interpreting the desired control behavior and effectively translating it into executable code, all while rigorously respecting operational constraints and predefined decision boundaries.

In this section, we formalize the approach using the canonical model, where each element is explicitly labeled as tok to clearly distinguish it from control-level system concepts. To structure the next-token generation process, we introduce an iteration counter $k$, which indexes each step. The state at step $k$, denoted by $S_k^{\text{tok}}$, primarily comprises the current sequence of tokens:

$$S_k^{\text{tok}} = (v_k^0, \ldots, v_k^j), \tag{4}$$

where $j$ represents the number of tokens generated by step $k$. Each token $v_k^j$ belongs to the discrete token space $\mathcal{V}_k$.[1] Typically, the input sequence $S_0^{\text{tok}}$ may contain an initial task description (e.g., "Implement a trading strategy for a battery storage") or encoded context about the target system, like the battery storage capacity. It may also include instructions that influence the behavior of the LLM itself, for example in the form of a system prompt.

Here the LLM functions as a policy $X_k^{\pi^{\text{LLM}}}$ at token level, mapping input $S_k^{\text{tok}}$ to a new token $x_k^{\text{tok}}$, similar to [15, 41]:

$$x_k^{\text{tok}} = X_k^{\pi^{\text{LLM}}}(S_k^{\text{tok}}). \tag{5}$$

Within an LLM, the policy comprises several architectural components, including embedding layers, transformer blocks, an output head, and a sampling mechanism, as detailed by Vaswani et al.[40]. Each of these components can be characterized by a specific function $f^{\text{LLM}} \in \mathcal{F}^{\text{LLM}}$ and associated parameters $\theta^{\text{LLM}} \in \Theta^{\text{LLM}}$. Ultimately, it is the interaction of these components that determines how the description of the high-level policy influences the next predicted token and is thus ultimately translated into executable code—and thus directly influences the resulting control behavior within the energy system.

The parameters of these components are primarily learned during pretraining, with potential refinement through subsequent posttraining stages [6] or runtime tuning [32]. Sampling strategies—such as greedy decoding or temperature-based sampling—are typically governed by a separate set of functions and parameters, enabling adaptive strategies during inference [7, 10, 47].

After sampling, the evolution of the token sequence at step $k + 1$ is governed by the sequence update mechanism $S^M$, defined as:

$$S_{k+1}^{\text{tok}} = S^{M,\text{tok}}(S_k, x_k^{\text{tok}}, W_{k+1}^{\text{tok}}).$$

During the inference stage, the function $S^{M,\text{tok}}$ incrementally extends the token sequence $S_k^{\text{tok}}$ by appending the newly generated token $x_k^{\text{tok}}$. Additionally, it incorporates any available exogenous information $W_{k+1}^{\text{tok}}$ at the appropriate position within the sequence. Such external inputs can result from so-called tool calling [17, 26], in which the inference is interrupted, external software tools are used for, e.g., querying energy market prices, and the results are fed back into the model. It is also possible to inject specific tokens [23] in order to control "thought processes" in a targeted manner.

At this point, we focus exclusively on the inference-time behavior of $S^{M,\text{tok}}$. The training dynamics—where the token sequence is typically reset or reconstructed at each step [48]—are not considered here. While our approach can, in principle, be extended analogously to the training setting, we restrict our analysis in this paper to the inference mode.

### 5.3.1 Objective Function.
Given an initial high-level policy description, represented as a token sequence $S_0^{\text{tok}}$ and enriched with contextual details—such as system constraints or energy-related objectives—the goal is to derive an optimal LLM-based policy specification, denoted by $\pi^{\text{LLM}} = \{f^{\text{LLM}}, \theta^{\text{LLM}}\}$. Over a total of $K + 1$ sequential generation steps, the corresponding policy $X^{\pi^{\text{LLM}}}$ is expected to generate a token sequence that explicitly encodes the intended control logic. After generating the final token and appending it to the sequence, the resulting token sequence $S_{K+1}^{\text{tok}}$ must fully represent the intended control policy $X^{\pi^{\text{es}}}$.

This task is formulated as an online optimization challenge, accommodating both initial model training (pretraining), potential posttraining refinements, and ongoing adaptation based on actionable feedback from the operational environment [37, 45, 54]. This adaptive approach supports autonomous task resolution within a self-play paradigm, continuously optimizing the policy parameters.

Furthermore, we include the scenario of discrete model selection for a given task $S_0^{\text{tok}}$, in which the most suitable pretrained LLM model is identified and (dynamically) selected from a candidate pool.

This process is coherently captured by the following objective function:

---
[1]For the sake of readability, we use the term 'token' to refer to both the token value and its corresponding token ID. Given the one-to-one mapping between the two, distinguishing them in notation is unnecessary for the purposes of this work.

$$\max_{\pi^{\text{LLM}}=\{f^{\text{LLM}},\theta^{\text{LLM}}\}} \mathbb{E}\left\{\sum_{k=0}^{K} C_k^{\text{tok}}\left(S_k^{\text{tok}}, X_k^{\pi^{\text{LLM}}}(S_k^{\text{tok}}), W_{k+1}^{\text{tok}}\right) \mid S_0^{\text{tok}}\right\}.$$

The cost function $C_k^{\text{tok}}$ evaluates how well the generated control policy adheres to the instructions and contextual constraints encoded in the initial token seed $S_0^{\text{tok}}$. These constraints include functional requirements such as capacity limits and operational bounds, as well as formal specifications like code formatting, interface definitions, or structural guidelines.

Furthermore, $W_{k+1}^{\text{tok}}$ is used to incorporate context information that may contain findings and restrictions, e.g., limit values for the feed-in of PV electricity according to online searches.

The final output sequence $S_{K+1}^{\text{tok}}$ may contain the intended control logic, but it can also include superfluous or incorrect elements, such as explanatory comments or syntax errors. Similarly, intermediate outputs $S_{k+1}^{\text{tok}}$, generated for code testing in a virtual sandbox during inference—such as with o3 [26]—may exhibit similar issues. To obtain a clean and executable control policy, we introduce a dedicated post-processing function:

$$X_t^{\pi^{\text{es}},k} = \Phi(S_{k+1}^{\text{tok}}).$$

Here, the index $k$ is included in the equation for completeness but will not be considered further in the paper to avoid confusion with other indices in the subsequent sections.

The post-processing function ensures that $X_t^{\pi^{\text{es}}}$ is an *explicit, code-based* policy function. In contrast, Wang et al. [42] propose a *direct-LLM-policy*, where the LLM itself would determine the actions $x_t^{\text{es}}$ at the control level, represented by $X^{\pi^{\text{es,LLM}}}$. In this case the decision logic *implicitly* remains encoded in the parameters of the LLM. However, the direct use of this direct-LLM-policy for control commands proves to be rather unsuitable, especially for use in the energy environment, as its actions are unpredictable and not reproducible—especially in safety critical context. By comparison, our explicit code policy representation is more transparent, auditable, and better suited for practical deployment.

So far, we have focused on translating predefined control policy statements into executable source code using LLMs. In the next chapter, we shift our attention to the origin of these statements, completing the hierarchical decision framework. We investigate how such high-level specifications are initially generated and what other decision mechanisms are available to guide the LLM's implementation behavior.

## 5.4 Level 3: Policy Search on Meta Level

As the highest level in the hierarchy, this layer governs the exploration and selection of control strategies. It leverages system-level feedback and performance metrics to adaptively guide policy search across function classes and policy types. A key task at this stage is to determine the degree of influence exerted on downstream code generation—ranging from explicit prompt instructions (e.g., "implement a linear program") to indirect modulation of model behavior (e.g., through temperature settings). This strategic control shapes how LLMs translate abstract intent into executable policies.

In both cases, the objective is to identify and iteratively refine the policy characteristic $\pi^{\text{es}}$ that optimizes the cost–benefit trade-off within the energy system.

### 5.4.1 Canonical Elements.
Each meta-level step is indexed by $i = 0, \ldots, I$ and represents a complete pass through the control policy creation pipeline—from high-level strategic design to low-level implementation. In online scenarios, for instance, a meta-decision may be triggered periodically, such as once per week. In offline settings, the frequency of meta-level steps may simply reflect how often the control policy generation is manually triggered on a local machine. In the following, we will list all canonical elements at this level with meta noted.

The corresponding states of the meta-state $S_i^{\text{meta}}$ include, on the one hand, elements directly derived from the energy system. These encompass, for example, historical time series such as heat consumption or evaluated metrics, such as the costs incurred by the executed control policy.

On the other hand, the meta-state may also contain information that provides insights into the quality of the generated code. This includes, for instance, indicators such as the number and severity of constraint violations in the implementation, or error messages that occur during execution.

Based on this state, the meta-policy function determines the next meta-action $x_i^{\text{meta}}$ as

$$x_i^{\text{meta}} = X_i^{\pi^{\text{meta}}}(S_i^{\text{meta}})$$

which modifies $\pi^{\text{LLM}}$ on the lower level for better policy generation. These adjustments typically involve two complementary ways of influencing:

(1) **LLM Input** ($S_0$ and/or $S_k$): The meta level can manipulate the input for the lower-level LLM by creating the prompt first or inserting predefined token sequences later during inference. These inputs may encode contextual information about the system, functional requirements, or corrective guidance, thereby steering the generation process. A variety of approaches exist for such lower-level LLM guidance, ranging from simple prompt templates to structured trees composed of predefined prompt snippets [35], up to more advanced setups such as a secondary LLM that dynamically creates prompts for the main model, or complex multi-agent systems where several LLM agents collaborate to solve tasks [38].

(2) **LLM Characteristics** ($\pi^{\text{LLM}}$): On the other hand, the meta level can also directly influence the internal configuration of the LLM and thus intentionally change its response behavior to a specific request. This includes, in particular, the dynamic adjustment of sampling parameters $\theta^{\text{LLM}}$—such as temperature tuning [53, 55]—or the adaptive modification of training weights, for instance through reinforcement learning based on environmental feedback, demonstrated by Surina et al. [37].

Both types of interventions aim to more effectively steer the LLM's behavior and to iteratively refine the resulting control policies $X_t^{\pi^{\text{es}},i}$ across multiple iterations $i$.

The meta-state evolves according to a transition function that integrates feedback from the previously executed meta-action and any resulting observations:

$$S_{i+1}^{\text{meta}} = S^{M,\text{meta}}(S_i^{\text{meta}}, x_i^{\text{meta}}, W_{i+1}^{\text{meta}}),$$

where changes in the meta-state are primarily driven by the outcomes of past meta-actions—such as accumulated performance metrics, updated LLM parameters or constraint violations observed during policy execution at the lower levels. Exogenous variables $W_{i+1}^{\text{meta}}$ refer to external information that only becomes available after a meta-decision has been made in the real world or a simulator. This includes feedback from the energy system (e.g., measured values, costs, constraint violations), context changes (e.g., market prices, weather data, user behavior) and technical information on code execution (e.g., errors, warnings, test results). This data flows into the next meta-state and is used for the evaluation and targeted adjustment of the LLM-based control strategy.

*Objective Function.* The overarching goal of the meta level is to identify a sequence of strategic decisions $x_i^{\text{meta}} = \{f^{\text{meta}}, \theta^{\text{meta}}\}$ that progressively enhances—directly or indirectly—the quality of the generated control policies. The corresponding optimization problem involves identifying a $\pi^{\text{meta}}$ that effectively influences the lower-level behavior of the LLM over the meta-horizon $I$, such that the expected cumulative contribution is maximized. This utility, denoted by $C_i^{\text{meta}}$, quantifies the effectiveness of the current control policy when deployed in the target environment. This could be, for example, the average costs or the runtime of the implemented control policy.

Formally, we define:

$$\max_{\pi^{\text{meta}}} \mathbb{E}\left\{\sum_{i=0}^{I} C_i^{\text{meta}}\left(S_i^{\text{meta}}, X^{\pi^{\text{meta}}}(S_i^{\text{meta}}), W_{i+1}^{\text{meta}}\right) \mid S_0^{\text{meta}}\right\}. \quad (6)$$

By formulating the optimization problem on the meta level as an online process, we explicitly account for the possibility of dynamic adjusting the meta-policy over time. Even when the underlying system model or LLM weights remain unchanged, $\pi^{\text{meta}}$ can evolve iteratively—based on observed feedback, performance metrics, or encountered constraint violations. This implies that, at the beginning of an investigation, simpler meta-search strategies may be employed—for example, a prompt template with a greedy character, where only the most recently best-performing control strategy is further refined. As more insights into the system's response behavior become available, the complexity of the approach can be gradually increased. For instance, the use of a second LLM to dynamically prompt the lower-level LLM can then be considered. This enables a more systematic exploration of the solution space toward a control policy that yields increasingly effective behavior within the energy system.

## 6 Implementation: Hierarchical Decision-Making with Agentic Policy Search

Having introduced the theoretical foundations of the hierarchical decision-making structure with its three levels—control, generation and meta—we now move on to its practical implementation. This chapter describes how the conceptual building blocks are transferred into a modular system architecture that enables autonomous decision-making and continuous optimization in energy technology systems. Figure 1 shows the functional structure of this architecture and illustrates the interaction of the three levels in the context of an integrated energy system. We follow the taxonomy of Sapkota et al. [33], in which AI agents are specialized, LLM-based systems, while agentic AI refers to coordinated, autonomous multi-agent systems with memory and task planning.

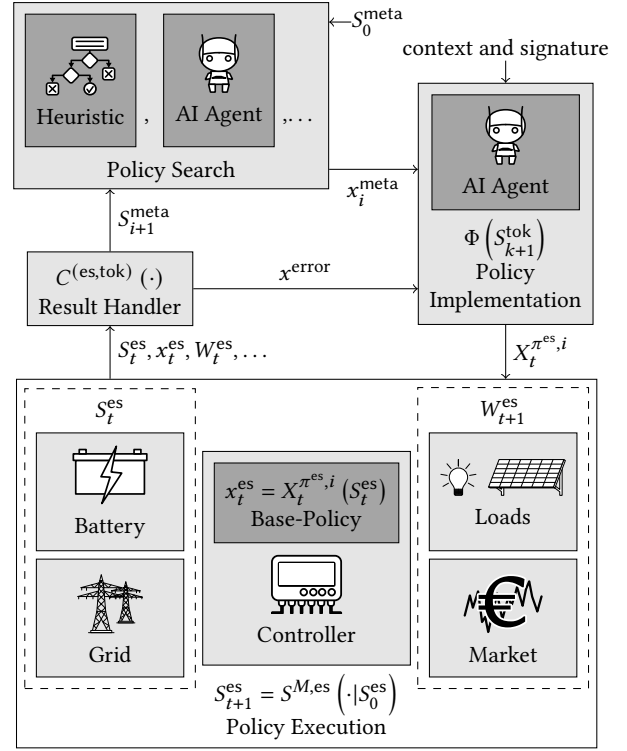

Fig. 1. System architecture of Agentic Policy Search (APS) framework with application in the energy domain

The APS-framework implements the multi-level decision-making structure described in Section 5. At the top level, the Meta-Search orchestrates the exploration across multiple control policies $X_t^{\pi^{\text{es}},i}$. This process is governed by one or more meta-policies, which act over multiple iterations $i$. These range from simple heuristics to Meta-LLMs, which dynamically generates the prompts for the lower-level LLM, or alternatively, mechanisms for tuning its parameters, e.g., through RL from environmental feedback. They can all be categorized in at least one of the main policy classes according to Powell [29] presented at the beginning (see Section 4).

In this context, LLMs can essentially be assigned to the class of PFAs, with the core components in the transformer architecture serving as a universal function approximator with tunable parameters, capable of representing any polynomial-time policy function with high accuracy when sufficiently scaled [20].

In alignment with this taxonomy and our architecture, we narrow the subsequent focus to the case, in which control policies are generated explicitly via prompting, rather than through implicit tuning of LLM-specific parameters $\theta^{\text{LLM}}$. Here, the meta-policy produces high-level textual descriptions that encapsulate the essential characteristics of the policy $\pi^{\text{es}}$.

Below that, the Policy Generator translates the abstract policy representation into an executable implementation $X_t^{\pi^{\text{es}}}$. To ensure compatibility with the target software environment, a formal software interface in the form of the class signature is provided as a constraint that must be satisfied during code generation. Initially, the LLM produces a complete textual output $S_{K+1}$, which is then processed in a post-processing step $\Phi$ to yield executable code. This step may involve removing extraneous elements or ensuring compliance with syntax rules. The generated code either fully encapsulates the functional logic, including parameters, or leverages existing software libraries (e.g., for solving linear programs). In the latter case, the availability of the respective library on the execution platform is a prerequisite for the correct policy deployment.

At the lowest level, the resulting control policy interacts either with the real energy system or its simulated representation. This can involve a real-world system, where the control policy is refined online over time based on environmental feedback, or a simulated environment that facilitates offline policy optimization. The system's transition function is determined either by physical or information-theoretic principles in reality or by the logic embedded in the simulation model.

The Result Handler evaluates policy execution outcomes within the energy system. The energy system's performance is quantified using an objective function $C^{(\text{es,tok})}(\cdot)$, containing both the contributions of implementation and execution level. Successful executions yield performance metrics, which are propagated back to the top level to inform subsequent policy refinement. These metrics may include energy-specific state variables (e.g., cost trajectories or decision histories) as well as runtime information. In case of execution errors, an error-handling routine is activated. This routine leverages feedback from caught runtime exceptions to iteratively correct faulty code segments, enabling autonomous policy repair. Details on implementation can be found in Algorithm 1 in Appendix A.

Unlike recent methods that integrate tool use (e.g., code execution, retrieval) directly into the LLM's reasoning chain within a monolithic design—such as ReTool [8] and ReSearch [2]—our framework employs a modular, hierarchical multi-agent architecture which can rather be assigned to the design patterns ReAct [51] and CodeAct [44]. By decoupling policy planning from implementation, it facilitates structured validation, modular development, and transparent verification. This stands in contrast to tightly coupled reasoning-action loops and supports reproducible, simulation-based evaluation of autonomous decision-making systems.

## 7 Use Case: Residential Energy System

Following the theoretical foundations, we now demonstrate the practical application of the framework in a simplified residential energy system. The goal is to illustrate the hierarchical interaction of control execution, code generation, and meta-level optimization in a realistic scenario. An LLM generates executable control code for the battery, which is iteratively refined based on simulation feedback.

### 7.1 Policy Execution: Model of a Residential Energy System

The simulated residential energy system consists of a battery, photovoltaic generation, household demand, a grid connection, and dynamic market conditions with fixed export and time-varying import prices. The control objective is to minimize total energy costs through optimal battery usage. In order to fulfill this goal, we are looking for a $\pi^{\text{es}}$ that provides the necessary decisions for the operation of the battery storage system. The system is formally defined using the five-element structure described in Section 4.

The entire simulation was implemented from scratch in python and unfolds over a finite time horizon $[0, T]$, discretized with a fixed time step $\Delta t$ resulting in the time steps $\mathcal{T} = \{0, 1, \ldots, T-1\}$. The corresponding parameters are given in Appendix B.

*7.1.1 State Variable.* The state variable $S_t$ at the beginning of time period $t$ encapsulates all information available to make a decision. For this residential energy system, the state is defined as a vector comprising the current energy level of the battery storage, the available power from the photovoltaic (PV) system, the energy demand of the household, and the prevailing market prices for electricity import and export.

$$S_t^{\text{es}} = (R_t^S, P_t^{\text{pv}}, L_t, p_t^{\text{buy}}, p_t^{\text{sell}}),$$

where:

- $R_t^S$ [kWh]: Battery state of charge at time $t$, representing the amount of stored energy.
- $P_t^{\text{pv}}$ [kW]: Power output of the photovoltaic system at time $t$.
- $L_t$ [kW]: Household electricity demand at time $t$.
- $p_t^{\text{buy}}$ [€/kWh]: Grid electricity price for energy imported at time $t$.
- $p_t^{\text{sell}}$ [€/kWh]: Feed-in tariff received for energy exported to the grid at time $t$.

*7.1.2 Decision Variable and Policy.* The decision variable $x_t^{\text{es}}$ represents the action taken by the control policy at time $t$ of the residential energy system simulation. The decision is determined on the LLM-generated control policy $X_t^{\text{es},\pi_i}$ at iteration step $i$. In this context, the decision is the power into or out of the battery:

$$x_t^{\text{es}} \in \mathcal{X}_t^{\text{es}} = [-D_{\max}, C_{\max}],$$

where $C_{\max}$ is the maximum charging power and $D_{\max}$ is the maximum discharging power, both in [kW]. A positive value $x_t > 0$ indicates charging the battery, while a negative value $x_t < 0$ indicates discharging the battery. The decision must respect the physical limits of the battery's power capacity. In cases where the control policy violates these constraints, the simulation environment enforces the admissible power limits by clipping values outside the range $[-D_{\max}, C_{\max}]$.

### 7.1.3 Exogenous Information.
The exogenous information $W^{es}_{t+1}$ represents uncertain events that occur between time $t$ and $t + 1$ and influence the state at $t+1$. In this model, $W^{es}_{t+1}$ drives the evolution of the PV generation, household demand, and market prices by distinct models:

- **PV generation** $P^{pv}_{t+1}$: Sinusoidal daily profile (06:00–18:00), multiplied by random noise and clipped at the nominal power of the PV module, taking solar irradiance into account,
- **Load profile** $L_{t+1}$: Typical household profile with base load, Gaussian morning peak (7:30) and evening peak (19:00), plus additive noise, and
- **Market prices** $p^{buy}_{t+1}/p^{sell}_{t+1}$: Buying price showing sinusoidal peaks (morning/evening) with random volatility, while feed-in tariff remain constant.

For each simulation run, we set a fixed random seed to ensure reproducibility of the stochastic time series, while preventing knowledge of future exogenous information during the simulation run for the control policy. A representative example of such a simulation run is provided in Appendix E.

### 7.1.4 Transition Function.
The transition function describes how the state evolves from $S^{es}_t$ to $S^{es}_{t+1}$. The primary dynamic element affected by the decision $x^{es}_t$ is the battery's state of charge. The battery state updates according to the following rule, accounting for charging and discharging efficiency $\eta$:

$$R^S_{t+1} = R^S_t + \eta \cdot \max(x_t, 0)\Delta t - \frac{1}{\eta} \cdot \max(-x_t, 0)\Delta t$$

subject to $0 \leq R^S_{t+1} \leq R^S_{max}$ with $R^S_{max}$ being the maximum storage capacity of the battery system. The other components of the state $(P^{pv}_{t+1}, L_{t+1}, p^{buy}_{t+1})$ transition based on the realization of exogenous information $W_{t+1}$. The feed-in tariff $p^{sell}_{t+1}$ is assumed to remain constant.

### 7.1.5 Objective Function.
The contribution function $C^{es}$ quantifies the cost incurred at time $t$ as a result of being in state $S_t$ and applying the decision $x_t$. The objective is to minimize the total energy cost. The net power imbalance at time $t$, considering PV generation, demand, and battery power flow, is given by:

$$g_t = L_t - P^{pv}_t - x_t.$$

This imbalance is resolved by interaction with the main grid. If $g_t > 0$, energy is imported; if $g_t < 0$, energy is exported. The cost for time step $t$ is then calculated based on the energy exchanged with the grid and the respective prices:

$$C^{es} = \left[\max(g_t, 0) \cdot p^{buy}_t - \max(-g_t, 0) \cdot p^{sell}_t\right] \Delta t.$$

This represents the cost of imported energy minus the revenue from exported energy during the time interval $\Delta t$.

The overall objective is to determine a control policy $\pi^{es}$ that selects actions at each time step $t$ of the simulation so as to maximize the total expected cumulative value of the energy system utility function $C^{es}$ over the finite horizon $T$ (see Section 5.2.2).

To solve this optimization problem, we employ the control policy defined by the code generated through the language model, which dynamically determines the system's behavior based on the current state.

### 7.1.6 Evaluation of implemented Control Policies.
To assess the quality of the generated control policies, we define a benchmark based on a globally optimal solution for $C^{es}$. This reference is computed by solving a deterministic linear program over the entire time horizon $T$, assuming perfect foresight of all exogenous variables (PV generation, household demand, and market prices).

The optimization minimizes total energy cost by determining optimal battery charging/discharging as well as grid import/export actions, subject to system constraints and energy balance conditions. For evaluation purposes, two variants of the benchmark are considered:

- A **finite-horizon optimum**, assuming a fixed initial battery state $R^S_0$, and
- A **steady-state optimum**, which enforces periodic boundary conditions $R^S_T = R^S_0$.

A detailed mathematical formulation of the benchmark model is provided in Appendix D.

### 7.1.7 Model Interface and Runtime Evaluation.
To enable interaction between the simulation environment and a control policy, a standardized interface was implemented. This interface exposes a predefined Python class structure with a `take_action` method, which is invoked at each simulation step. The method receives the current system state (battery level, PV generation, demand, market prices, and capacity) as input and returns the control action for the battery. The interface is embedded into the LLM prompt to ensure that the generated code adheres to the simulation requirements. A complete code listing and further implementation details are provided in Appendix C. Additionally, the runtime for each policy execution is recorded, enabling the evaluation of policy performance over successive iterations.

## 7.2 Policy Implementation: LLM generates Controller Code

This level represents the direct generation of executable control code for energy systems. It builds upon task descriptions and contextual guidance received from the meta-policy, enabling the generation of complete Python implementations matching the python signature in Appendix C. The objective here is to translate abstract goals into functional, rule-compliant generative behavior.

### 7.2.1 Decision Variable and LLM-Policy.
For our policy $X^{\pi^{LLM}}$ we choose a pretrained LLM model, where the characteristics of the token generation are determined by the architecture and the weights tuned during pre- and post-training. The decision $x^{tok}_k$ refers to the token at generation step $k$, resulting in the control policy code, using $\Phi$ (see Section 5.3.1). For simplicity, we choose the option of using the final token sequence $K$ for code generation, without intermediate code evaluation during inference. The selection process of a suitable LLM will be further discussed in Section 7.2.4.

### 7.2.2 State Variables and Contextual Instructions.
At each overall meta-iteration $i$, the generation-level LLM observes a structured input comprising:

- **Context**: High-level information outlining the overall objective and guiding principles that shape token generation behavior.
- **Task Description** (task_description): An instruction, refined or generated by the meta-policy, defining the function type $f^{es}$ and parameter $\theta^{es}$ of the new control policy.
- **Function Signature** (policy_signature): The precise structure (see Appendix C) that the generated implementation must follow, ensuring syntactic and functional compliance within the simulation framework.
- **System Constraints**: Domain-specific boundary conditions such as battery capacities, energy balance requirements, and technical limits that guide valid behavior.

These elements form the full initial input state $S_0$ of the prompt as presented in the following:[2]

```
You are an expert Python developer.

Develop an intelligent battery management policy to
    optimize energy costs while satisfying the
    demand_sequence.
The policy must make strategic decisions about:
1. When to charge the battery (buy & store energy)
2. When to discharge the battery (sell & discharge
    energy)
3. When to directly purchase from the market

Key Constraints:
1. Battery Capacity:
   - 0 ≤ energy_stored ≤ max_energy_stored
   - Battery charge must stay within physical limits
   - power_discharge ≥ -5
   - power_charge ≤ 10

2. Energy Conservation:
   - energy_discharged ≤ energy_stored
   - Cannot discharge more energy than stored

3. Demand Coverage:
   - power_bought + power_discharged ≥ power_own_demand
   - Must meet energy demand_sequence in each timestep

Structure example:
{policy_signature}

Implementation instructions:
{task_description}

Provide the final implementation without Markdown
    formatting or additional comments outside the class.
```

In case of failures (e.g., due to simulation errors, missing constraints, or syntax issues), a fallback routine is being activated, which formulates a repair prompt based on diagnostic feedback error_message about erroneous policy_code.

```
You are an expert Python developer debugging a battery
    management system implementation.

A BatteryPolicy implementation has failed in the
    simulation environment with the following:
Error Message:
{error_message}

Failed Code:
```python
{policy_code}
```
Task:
Fix the implementation errors while maintaining the
    original strategy where appropriate.

Expected output structure:
{policy_signature}

Return only the corrected Policy class implementation
    without markdown formatting or extra comments outside
    the class.
```

If the generated control policy code is still incorrect after 5 failed attempts, the iteration loop $i$ is restarted.

*7.2.3 Exogenous Information and Transition Function.* We assume that the selected LLM operates deterministically based on input prompts and internal model state. During inference, no exogenous dynamic information $W_{t+1}^{tok}$ is ingested into the token sequence $S_k^{tok}$. Therefore, all relevant signals are encoded in the input prompt $S_0^{tok}$ itself, including constraints and goals.

The state transition $S_k^{M,tok}$ on policy level is implicitly defined by the next-token prediction mechanism of the LLM. The system evolves based on the input prompt and the model's internal distribution over possible token continuations.

*7.2.4 Cost Function and Optimization Objective.* We are looking for the policy (here: choice of an LLM) that delivers the best result after a limited number of attempts. The selection of a pretrained model is formulated as an optimization problem in Section 5.3.1 with the goal of (1) minimizing the semantic deviation between prompt statement and model response, (2) strictly adhering to formal constraints such as interface specifications and (3) keeping the inference costs low. These objectives are embedded into $C^{tok}$. At this point, we will stick to a more abstract description before we delve deeper into the optimization problem in the following chapter.

Since analytical modeling of these target variables is difficult, the solution is empirical: By benchmarking several LLMs on software-related tasks, quality and costs are systematically compared in order to select the optimal model under given requirements.

Gemini 2.5 Flash was selected because it offers a convincing balance between performance and cost. In the LiveCodeBench v5 benchmark, it achieved one of the highest score of all tested models for code generation. It also performs well in the MRCR benchmark for long contexts and the AIME 2025, which measures mathematical capabilities. At the same time, it is one of the cheapest models among competitors [11]. This combination makes it a suitable choice for efficient, high-quality code generation from a cost perspective. Further information such as model specifications or the costs of the generated tokens can be found in Appendix F.

However, it is important to consider that selecting this closed-source model, hosted on external servers, limits our ability to influence the token generation process during inference, particularly through advanced techniques such as token injection.

### 7.3 Meta Search: LLM searches for the best Control Policy

Building on top of the generation layer, the meta level generates the instructions that should be followed when implementing a new control policy. This instruction is the task description variable that was previously embedded in the lower level LLM prompt. In the following, we will focus on the creation of this statement.

---
[2]Variables are given in curly brackets. Some variable names have been slightly renamed for reasons of clarity.

*7.3.1 Decision Variable and Meta-Policy.* The meta-policy $X^{\pi^{\text{meta}}}$ here in our use case consists of a second pretrained LLM combined with a distinct myopic policy generating discrete prompts in each iteration $i$. This meta-LLM generates dynamically the content of the variable `task_description` for the lower level—code generating—LLM. It employs a simple greedy rule: whenever a new candidate achieves lower total cost than previous attempts, this approach is further refined. Otherwise, a new one is chosen. Additionally, the policy tracks performance *stagnation* and automatically switches to exploration when no significant improvement is detected over multiple iterations.

*7.3.2 State Variables and System Metrics.* In our simulation several state variables $S_i^{\text{meta}}$ from the environment are monitored at each step $i$. These metrics allow a high-level performance assessment.

Notably, no explicit references to the time series of $S_t^{\text{meta}}$ or $W_{t+1}^{\text{meta}}$ (e.g., PV generation, load profiles, market prices) are being provided to the meta-policy. Instead, such external factors are implicitly represented through their aggregated impact on performance metrics, such as:

- **Total cost of last simulation run** $c_i^{\text{sim}}$ (`total_cost`): The total cumulative cost during the most recent simulation run,
- **Best cost achieved so far** (`best_cost`): The lowest total cost found across all iterations $i$ of the meta-search process,
- **Iteration count** (`iteration_count`): The number of iterations $i$ completed so far,
- **Battery utilization** (`utilization`): The percentage of time the battery was actively charging or discharging during the simulation,
- **Average state of charge** (`avg_soc`): The average energy level of the battery across the entire simulation horizon, and
- **Price volatility** (`price_volatility`): The ratio of the standard deviation to the mean of electricity prices, indicating price fluctuations.
- **Cost history** (`cost_history`): The last costs of the last tries.

This abstraction helps to reduce input space complexity and prevent reward hacking, meaning the model is prevented from exploiting periodic patterns in exogenous inputs or the determinism of a fixed random seed to gain an unfair predictive advantage.

The prompt $S_0^{\text{meta}}$ for Meta-Policy consists of different subsections. The main prompt includes both the individual system states and the reference Python code from the previous iteration stage. Based on this, the task is to analyze the most recent implementation approach and either refine it further or pursue an alternative method—depending on the exploration strategy and task mode.

```
You are an expert developing an intelligent battery
    management system.
Current Total Cost: {total_cost}
Best Cost Achieved: {best_cost}
Iteration: {iteration_count}
Battery Utilization: {utilization}%
Average State of Charge: {avg_soc}
Price Volatility: {price_volatility}
Performance History (last 5 costs): {cost_history}
```

```
Current Implementation:
```python
{policy_code}
```

{explore_or_refine_instruction}

Your task:

1. Analyze the current implementation's strengths and
    limitations
2. {task_mode}
3. Provide specific parameter values and implementation
    details
4. Explain expected impact on cost and system behavior

Focus on CONCRETE improvements that can be implemented
    immediately.
```

The exploration strategy `explore_or_refine_instruction` may either include fine-tuning the approach

```
Suggest ONE specific improvement to the existing approach
```

or pursue a different method.

```
Propose a novel approach that fundamentally rethinks how
    we make charging/discharging decisions
```

Here, the `task_mode` provides additional contextual guidance on whether the current policy should be refined

```
The current approach shows potential. Focus on targeted
    improvements while maintaining core strategy.
```

or whether a new strategy should be pursued instead.

```
The current approach shows stagnation. Consider a
    fundamentally different strategy to optimize the
    battery management system.
```

*7.3.3 Exogenous Information and Transition Function.* The simulation produces $W_{t+1}^{\text{meta}}$, which are used to evaluate the effectiveness of energy management strategies. These quantities correspond to the previously introduced state variables $S_i^{\text{meta}}$, which are computed within the simulation environment. Accordingly, the transition function $S_i^{M,\text{meta}}(\cdot)$ represents a simple update function for the meta-states. As outlined previously, we deliberately avoid passing raw time series directly to the meta-policy in order to prevent reward hacking and overfitting to specific patterns.

*7.3.4 Cost Function and Optimization Problem.* We aim to identify the best meta policy—here, the choice of a pretrained LLM $\pi^{\text{meta}} = \{f^{\text{meta}}, \theta^{\text{meta}}\}$—that achieves the lowest observed cost $c_i^{\text{sim}}$ within a limited number of $I+1$ iterations. The model selection described by Equation (6) can be reformulated as a finite-horizon optimization problem, following the structure in [29]:

$$\max_{\pi^{\text{meta}}} \mathbb{E}\left\{ F\left(x^{\pi^{\text{meta}},I}, W^{\text{meta}} \mid S^{0,\text{meta}}\right) \right\},$$

where:

- $S^{\text{meta},0}$ – initial prompt from Section 7.3.1 at the meta level
- $x^{\pi^{\text{meta}}}$ – token sequence produced by the selected model $\pi^{\text{meta}}$ after $I$ Iterations, in particular the generated variable `task_description`

- $W^{\text{meta}}$ – metrics from simulation environment, such as `total_cost` or `avg_soc`
- $F(\cdot) = C^{\text{meta}}$ – contribution function at the meta level, e.g., defined as the negative simulation cost $C^{\text{meta}} = -c_i^{\text{sim}}$.

In practice, we solve this problem by benchmarking several LLMs and selecting the one with the highest empirical performance. Following the evaluation described in Section 7.2.4, we choose *Gemini 2.5 Flash Preview* as the second-level LLM with the same model specification listed in Appendix F. This choice reflects a favorable trade-off between inference cost and performance, while achieving state-of-the-art results across multiple benchmarks [11].

## 8 Results

The evaluation is based on 20 episodes, each of which was carried out over ten iterations $i$ of a simulation cycle across the time interval $[0, T]$. [3] The aggregated representation (Figure 2) reveals key trends in the cost trajectories across all iterations.

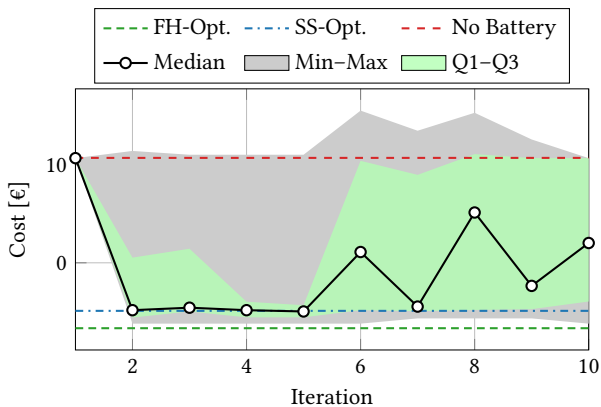

Fig. 2. Evolution of cumulative cost over iterations across all runs. The black line shows the median cost at each iteration, the green shaded area marks the interquartile range (25th–75th percentile), and the gray band spans from the minimum to the maximum observed cost. The dashed green line (−6.67 €), the dashed blue line (−5.20 €), and the dashed red line (10.7 €) indicate the benchmark baselines obtained by the finite-horizon optimum ("FH-Opt."), the steady-state optimum ("SS-Opt."), and the no-battery strategies, respectively.

An overall reduction in operational costs is already evident in the early stages compared to the no-battery reference scenario (10.70 €), which assumes no active battery dispatch. From iteration 2 onward, the median cost gradually declines below zero, which means that a profit is being made. Notably, iteration 5 marks a particularly stable phase: the median reaches a local minimum, and the interquartile range (Q1–Q3) is at its narrowest, indicating high consistency among the generated control policies. During this phase, many runs converge towards the benchmark values achieved through explicit optimization—finite-horizon optimization (−6.67 €) and steady-state optimization (−5.20 €). This saturation effect after several rounds of iterative refinement is consistent with the findings reported by Wang et al. [43]. In later iterations—especially from iteration 6 onward—the variance in performance increases. This is also reflected in the runtimes, which show higher scores across different episodes from this point onwards (see Appendix G). The range between minimum and maximum widens, and the interquartile range expands, reflecting greater divergence in policy effectiveness.

---
[3]Episode 4, iteration 7 (41.8 €) and episode 16, iteration 10 (27.8 €) were identified as outliers and removed during post-processing.

The 200 generated control policies cover a wide range of representatives from three of the four policy classes previously introduced in Section 4: policy function approximations (PFAs), value function approximations (VFAs) and direct lookahead approximations (DLAs).

The most frequently represented form are PFAs in the form of rule-based strategies, which manage the charging and discharging behavior based upon, e.g., the current market price. Particularly noteworthy are the cases in which the generated control policy initializes an online parameter tuning with collected data during runtime in order to adapt parameters, such as threshold values. In addition, several approaches incorporate on-the-fly tuning mechanisms to improve the prediction of exogenous factors, such as price signals or photovoltaic yields.

By contrast, advanced approaches such as MPC—a member of the DLA class—were particularly error-prone because crucial software libraries were missing, forcing a restart after five unsuccessful attempts. The same practical limitations also hindered reinforcement-learning algorithms such as PPO [34] and DQN [22], both of which are related to the VFA class. The approaches from the class of cost function approximations (CFAs) were not represented at all among the implemented control policies.

Among the best control strategies were mostly representatives of PFA in the form of rule-based approaches. An exemplary simulation run (episode 16, iteration 5), illustrated in Figure 3, shows the temporal evolution of buy and sell actions as well as the corresponding battery state of charge. This control strategy first prioritizes self-consumption and then leverages dynamic price arbitrage: when grid prices are low, the system opportunistically charges the battery, and when prices are high, it discharges energy to the grid—always within the technical constraints of the system and guided by a historical price window spanning the last estimated 24 hours. The decision logic is greedy in nature, reacting solely to current conditions and past prices without forecasting future values. The state of charge fluctuates dynamically across the entire bandwidth of the battery's capacity. Within the charging and discharging boundaries, the action patterns exhibit high variability over time, primarily due to the cyclic nature of the underlying time series. This results in an almost optimal costs of up to −6.20 € compared to the no-battery scenario, indicating a net monetary gain over the analysis period.

## 9 Discussion

The results presented in this study demonstrate that the proposed APS framework effectively generates a near-optimal battery control policy within only a few iterations, despite lacking prior knowledge of future energy prices or photovoltaic generation. This rapid development is notable given that the process operates under incomplete information and minimal contextual input. By separating high-level strategy exploration (meta-policy search) and low-level implementation (explicit Python code generation), the framework enables a manageable and efficient optimization pipeline. The transparent and auditable nature of the executable code simplifies debugging and ensures robustness, as runtime errors and constraint violations can be efficiently isolated and resolved without disrupting the overall search process.

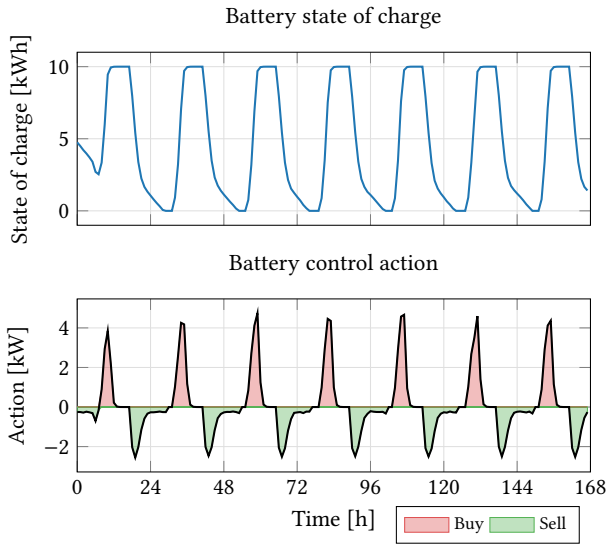

Fig. 3. Battery state of charge and corresponding control actions.

However, the increased variability observed in later iterations highlights intrinsic limitations related to robustness and scalability in policy development. A primary factor is the minimalistic prompt design, which currently provides insufficient structural, operational, and contextual details about the energy system, posing challenges for scalability to more complex scenarios. Additionally, the current meta-policy employs a myopic selection policy based solely on immediate cost performance, neglecting cumulative performance over time. This approach may favor short-term gains while ignoring longer-term stability. Once a control policy's improvement potential is exhausted—often visible as a performance plateau—further adjustments tend to degrade quality, increasing both costs and runtimes. A monitoring system would need to detect such saturation effects and, guided by a meta-policy, deliberately explore alternative strategies—such as representatives of the CFA policy class—without prematurely reverting to the saturated but previously successful approach at the first sign of performance decline.

The absence of a dedicated tuning phase also impacts the effectiveness of data-driven approaches, as internal policy parameters remain suboptimal. Furthermore, this study's exclusive reliance on simulation data omits critical real-world complexities, such as non-cyclical demand patterns, reliability concerns, and safety requirements. Additionally, outlier removal, while statistically justified, may obscure valuable insights into edge-case behaviors and potential failures.

For real-world deployment, several additional safeguards are essential to ensure safety, reliability, and long-term performance. Firstly, fail-safe mechanisms should activate a conservative fallback strategy whenever control policies exhibit faulty behavior or violate safety constraints. Secondly, high-resolution digital twins can rigorously validate policies prior to deployment, exposing vulnerabilities under edge-case scenarios such as unexpected demand spikes or grid disturbances. Thirdly, scheduled re-tuning intervals informed by field data help counteract performance drift resulting from structural system changes or external market developments. Finally, compliance with grid codes, hardened cybersecurity, and seamless integration into existing energy management systems are critical prerequisites for transitioning the proposed framework from a promising prototype to an industry-grade control solution.

## 10 Conclusion

This paper presents a hierarchical Agentic Policy Search (APS) framework for energy systems, integrating LLMs into a multi-layer architecture to automate the generation, evaluation, and refinement of executable control strategies. The approach effectively combines high-level strategic adjustments (meta-policy) with detailed, transparent software implementation, enabling continuous improvement and adaptation.

Applied to a residential energy system involving battery storage and photovoltaic generation, APS rapidly achieves substantial cost reductions. Median energy costs fall below the base scenarios without battery storage at an early stage, and individual runs almost reach the lower bound of a finite-horizon optimum. The explicit representation of control strategies as auditable and debuggable code offers clear advantages over opaque end-to-end LLM controllers, enhancing interpretability and reliability.

Nonetheless, experimental findings also reveal limitations, notably increased performance variability after initial successes, due primarily to short-sighted selection criterion. Addressing these issues is crucial for scaling APS to more complex and realistic environments.

Future research should enhance meta-policy mechanisms to balance short-term performance with long-term robustness and incorporate a fine-tuning phase to optimize data-driven strategies. Improved prompt design and richer environmental context are essential for effectively handling more sophisticated real-world scenarios. Additionally, integrating mechanisms to ensure security, reliability, and compliance will be critical for real-world deployments.

While initially demonstrated in energy systems, APS principles can extend broadly across other domains, especially as more reliable and powerful language models emerge, promising significant academic and industrial impacts.

## A  Algorithm

The procedure within the hierarchical decision structure can be concisely presented as shown in Algorithm 1. The algorithm first initializes the states for the energy system, the token series, and the meta-control layer. In each meta step (for $i > 0$), a meta-policy generates a high-level command $x_i^{\text{meta}}$ that serves as the initial state for the autoregressive token generation, where an LLM generates tokens over $K + 1$ steps and updates the token state accordingly. The final token state $S_{K+1}^{\text{tok},i}$ is transformed via $\Phi$ into a control policy for the energy system, which is then executed over multiple time steps to take concrete actions and update the energy system state with gathered observations. The resulting performance metrics are used to update the meta state, allowing the hierarchical decision-making and learning process to iteratively refine its strategies.

**Algorithm 1** Sequence of the hierarchical decision structure via agentic policy search

1: Initialize $S_0^{\text{tok},0}, S_0^{\text{es}}, X_t^{\pi^{\text{es}},0}$
2: **for** each meta step $i = 0, 1, \ldots, I$ **do**
3:    **if** $i > 0$ **then**
4:       $x_i^{\text{meta}} \leftarrow X_i^{\pi^{\text{meta}}}(S_i^{\text{meta}})$
5:       $S_0^{\text{tok},i} \leftarrow x_i^{\text{meta}}$
6:    **for** each token step $k = 0, 1, \ldots, K$ **do**
7:       $x_k^{\text{tok}} \leftarrow X^{\pi^{\text{LLM}}}(S_k^{\text{tok},i})$
8:       Observe $W_{k+1}^{\text{tok}}$
9:       $S_{k+1}^{\text{tok},i} \leftarrow S^{M,\text{tok}}(S_k^{\text{tok}}, x_k^{\text{tok}}, W_{k+1}^{\text{tok}})$
10:    $X_t^{\pi^{\text{es}},i} \leftarrow \Phi(S_{K+1}^{\text{tok},i})$
11:    **for** each time step $t = t^i, \ldots, T^i$ **do**
12:       $x_t \leftarrow X_t^{\pi^{\text{es}},i}(S_t^{\text{es}})$
13:       Observe $W_{t+1}^{\text{es}}$
14:       $S_{t+1}^{\text{es}} \leftarrow S^{M,\text{es}}(S_t^{\text{es}}, x_t^{\text{es}}, W_{t+1}^{\text{es}})$
15:    Update $S_{i+1}^{\text{meta}}$ with new metrics

## B  Residential Energy System Parameter

Table 1 summarizes the technical and economic parameters of the simulated residential energy system, including battery specs, load profiles, PV capacity, and electricity pricing. The simulation runs for 7 days with a 60-minute resolution, using a fixed random seed for reproducibility.

Table 1. Simulation parameters of the residential energy system scenario

| Parameter | Unit | Value |
| --- | --- | --- |
| Battery capacity | [kWh] | 10.0 |
| Battery round-trip efficiency | [%] | 90 |
| Initial state of charge | [%] | 50 |
| Base load demand | [kW] | 0.25 |
| Morning peak demand | [kW] | 1.25 |
| Evening peak demand | [kW] | 2.25 |
| PV nominal power | [kWp] | 5.0 |
| Simulation duration | [days] | 7 |
| Grid electricity price | [€/kWh] | 0.35 |
| Feed-in tariff | [€/kWh] | 0.08 |
| Price volatility | [%] | 10 |
| Simulation time step | [min] | 60 |
| Random seed | [-] | 42 |

## C  Model Interface

The following Python class `Policy` defines the function signature that will be invoked by the environment during the simulation runtime process. This signature is provided as a reference in the LLM's prompt to implement the `take_action` method.

```
class Policy:
  def __init__(self):
    """Initializes the policy.
    Parameters or internal states can be defined here
    """
    pass

  def take_action(self,
    # energy stored in the battery [\unit{\kWh}]
    current_energy_stored_kwh: float,
    # PV power generation [\unit{\kW}]
    current_pv_generation_kw: float,
    # household power demand [\unit{\kW}]
    current_demand_kw: float,
    # grid purchase price [euro/kWh]
    current_grid_buy_price: float,
    # grid feed-in tariff (sell price) [euro/kWh]
    current_grid_sell_price: float,
    # Maximum battery capacity [\unit{\kWh}]
    battery_capacity_kwh: float,
    ) -> float:
    """Determines the target action for the battery based
       on the current state.

    Returns:
      float: The target power for the battery [\unit{\kW
    }]
        positive: charging; negative: discharging;
        zero: no action
    """
    # --- Implement your logic here ---
    # Example: Always return 0 (no action)
    action_kw = 0.0

    # Return the calculated action
    return action_kw
```

The take_action method is designed to be called with the current state parameters, allowing the LLM to compute and return the optimal energy trading action. Additionally, class attributes can be used—by decision of the LLM—to define parameters such as thresholds or to store intermediate results from previous time steps. These stored values may support interim data analysis during runtime or even be used by an embedded lookahead models to improve future decision-making.

### D  Benchmark

We choose a global optimal benchmark obtained by formulating and solving a deterministic linear optimization problem over a finite horizon $T$ with time steps $\mathcal{T} = \{0, 1, \ldots, T-1\}$. Specifically, all exogenous time series (PV generation, load demand, and market prices) are fixed in advance for benchmarking purposes.

We introduce separate variables for charging $c_t \geq 0$ and discharging $d_t \geq 0$ powers of the battery storage at each time step $t \in \mathcal{T}$, with:

$$0 \leq c_t \leq C_{\max}, \quad 0 \leq d_t \leq D_{\max}$$

Furthermore, let $i_t \geq 0$ and $e_t \geq 0$ denote grid import and export powers, respectively. The state of charge of the battery $R_t^S \in [0, R_{\max}^S]$ evolves according to the transition function:

$$S_{t+1}^S = R_t^S + \eta \cdot c_t \Delta t - \frac{1}{\eta} \cdot d_t \Delta t,$$

subject to initial condition $R_0^S$.

The power balance at each time step $t$ is enforced by:

$$P_t^{\text{pv}} + d_t + i_t = L_t + c_t + e_t.$$

The objective is to minimize the total energy cost over the finite horizon:

$$\min_{c_t, d_t, i_t, e_t} \sum_{t=0}^{T-1} \left( i_t \cdot p_t^{\text{buy}} - e_t \cdot p_t^{\text{sell}} \right) \Delta t.$$

The solution of this optimization yields a globally optimal control sequence $\{c_t, d_t, i_t, e_t\}_{t=0}^{T-1}$. For comprehensive evaluation, two benchmarks are reported:

- **Finite-horizon optimum:** obtained by solving the optimization problem for a fixed initial condition $R_0^S$ and
- **Steady-state optimum:** enforcing cyclic boundary conditions, $R_T^S = R_0^S$, to evaluate the control policy in a steady-state (periodic) scenario.

### E  Exogenous Information

Figure 4 shows an example for the time series of electricity demand alongside on-site generation, illustrating the mismatches between consumption and local production, whereas Figure 5 presents an example for the evolution of the market buy price and the fixed sell price over time.

### F  LLM Model

The specifications presented in the following describe the interface of *Gemini 2.5 Flash Preview* as available via the OpenRouter API [27] at the time of publication. In addition, the number of tokens and the resulting costs of the experiment conducted are listed in Table 2. The data refers to both the LLM at level 2, which implements

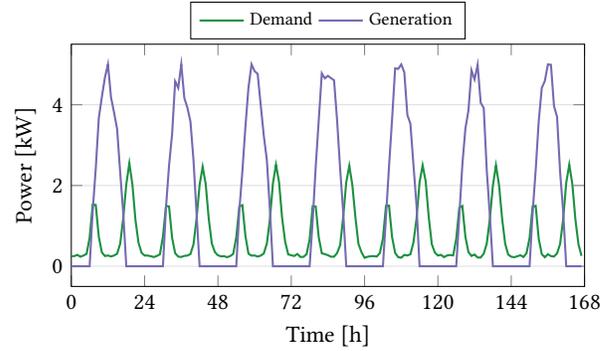

Fig. 4. Time series of electricity demand and on-site generation.

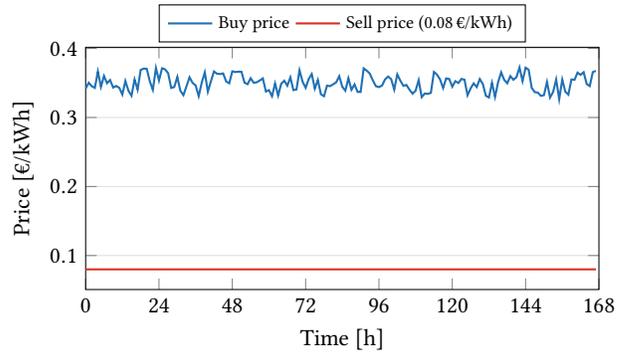

Fig. 5. Market buy price and fixed sell price over time.

Table 2. Technical Specifications of the Gemini 2.5 Flash Preview Interface via the OpenRouter API

| Parameter | Unit | Gemini 2.5 Flash Preview |
| --- | --- | --- |
| Model version | - | 04-17 |
| Context size | [tok] | 1.05M |
| Maximum output size | [tok] | 66K |
| Input cost | [$/M tok] | 0.15 |
| Output cost | [$/M tok] | 0.60 |
| Latency | [sec] | 0.65 |
| Avg. throughput | [tok/sec] | 91.85 |
| Temperature | - | 1 |
| Top-p | - | 0.95 |
| Top-k | - | 64 |
| Candidate count | - | 1 |
| Input tokens fed | [tok] | 5.31M |
| Output tokens generated | [tok] | 3.29M |
| Total cost | [$] | 2.77 |
| Total compute time | [h] | 6.19 |

the code, and the LLM at level 3, which performs the search across different control policies.

## G  Simulation Runtime

Figure 6 presents a heatmap where rows represent independent simulation episodes (1–20) and columns represent iterations within each episode (1–10). The color scale indicates the runtime per iteration in seconds, with brighter colors denoting longer durations.

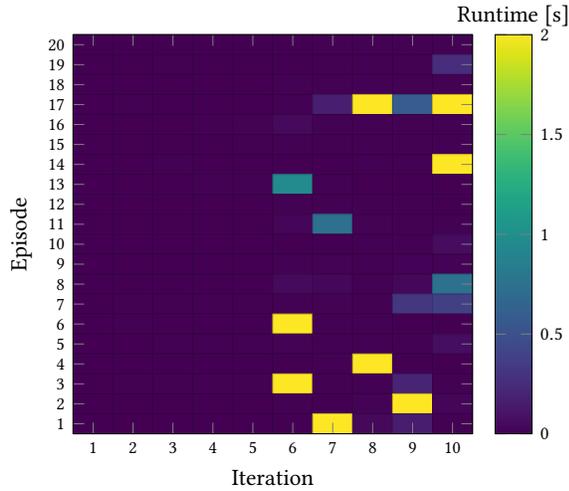

Fig. 6.  Runtime heatmap per iteration and episode.

Runtimes above 2 seconds are clipped for visualization. Although episodes run independently, a clear trend of increasing runtime after iteration 5 is visible. This is due to the growing computational load from repeatedly executing the forecast routine—including weighted averages and linear trend estimation—at every simulation step. The longest runtime appears in episode 4, iteration 10, caused by a particularly intensive forecast routine with parameter tuning applied at every simulation step.